\documentclass[journal,twoside,web]{ieeecolor}
\usepackage{generic}
\usepackage{cite}
\usepackage{amsmath,amssymb,amsfonts}
\usepackage{algorithmic}
\usepackage{graphicx}
\usepackage{textcomp}
\usepackage{float}
\usepackage[export]{adjustbox}
\usepackage[caption=false,font=footnotesize]{subfig}

\usepackage{etoolbox}
\makeatletter
\@ifundefined{color@begingroup}%
  {\let\color@begingroup\relax
   \let\color@endgroup\relax}{}%
\def\fix@ieeecolor@hbox#1{%
  \hbox{\color@begingroup#1\color@endgroup}}
\patchcmd\@makecaption{\hbox}{\fix@ieeecolor@hbox}{}{\FAILED}
\patchcmd\@makecaption{\hbox}{\fix@ieeecolor@hbox}{}{\FAILED}
\makeatother

\begin{document}
\title{Soft Error Susceptibility of 4F\textsuperscript{2} DRAM and FeRAM: Roles of Cell Structure and Storage Mechanism }
\author{Rudra Biswas, \IEEEmembership{Graduate Student Member, IEEE}, Jiahui Duan, \IEEEmembership{Member, IEEE}, Rajiv Joshi, \IEEEmembership{Fellow, IEEE}, Kai Ni, \IEEEmembership{Member, IEEE}, and Vijaykrishnan Narayanan, \IEEEmembership{Fellow, IEEE}
\thanks{This material is based upon work supported by the center for 3D Ferroelectric Microelectronics and Manufacturing (3DFeM2), an Energy Frontier Research Center funded by the U.S. Department of Energy, Office of Science, Office of Basic Energy Sciences Energy Frontier Research Centers program under Award Number  DE-SC0021118. }
\thanks{Rudra Biswas and Vijaykrishnan Narayanan are with the Department of
Electrical and Computer Engineering, Georgia Institute of Technology, Atlanta, GA, USA (e-mail: rbiswas45@gatech.edu). }
\thanks{Jiahui Duan, and Kai Ni are with the Department of Electrical
Engineering, University of Notre Dame, Notre Dame, IN 46556 USA.}
\thanks{Rajiv Joshi is with the IBM Thomas J. Watson Research Center,
Yorktown Heights, NY 10598 USA.}}

\maketitle

\begin{abstract}
Aggressive density scaling toward $4F^2$ vertical memory architectures introduces new radiation-reliability challenges, particularly as emerging memory technologies adopt fundamentally different storage and sensing mechanisms. This work systematically investigates heavy-ion-induced single-event effects in conventional 1T–1C or capacitorless 2T–0C DRAM, and single- and multi-capacitor ferroelectric RAM using experimentally calibrated 3D TCAD and mixed-mode simulations. Strike-induced floating-body effects are identified as a key mechanism enabling unintended charge transfer under nonzero drain–source bias. Our study finds that the density advantage of capacitorless DRAM comes at the expense of radiation resilience, exhibiting approximately an order-of-magnitude lower linear energy transfer upset threshold than 1T–1C DRAM (sub-1 versus $\leq$6 MeV·cm²/mg). While polarization-based storage in FeRAM prevents upsets at values up to 200 MeV·cm²/mg under zero-bias hold, half-selected cell operations can introduce vulnerabilities that reduce this threshold to 15 MeV·cm²/mg. Material properties, operating biases, and capacitor geometry are identified as key design parameters governing polarization disturbance. Furthermore, vertical ferroelectric capacitor stacking simultaneously improves memory density and radiation resilience by increasing the effective storage-node capacitance. These findings establish device- and architecture-level guidelines for jointly optimizing memory density and radiation resilience.

\end{abstract}


\section{Introduction}

\begin{figure}[t]
\centerline{\includegraphics[width=1.0\columnwidth]{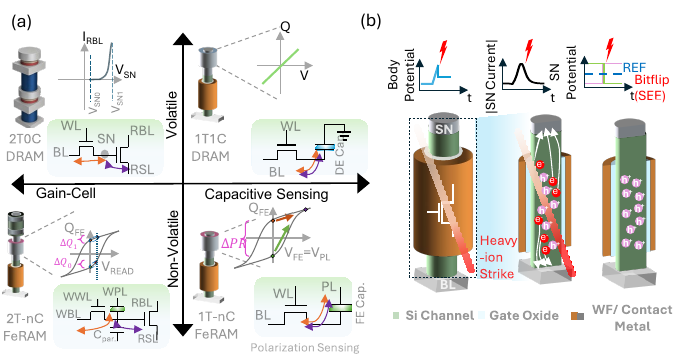}}
\caption{(a) Memory technology variations in storage mechanisms and cell architectures.
(b) Heavy-ion strikes in 4F² memory transistors can induce transient floating-body effects, leading to storage-node charge disturbances and single-event upsets.}

\vspace{-7mm}
\label{fig:intro}
\end{figure}

\IEEEPARstart{T}{he} rapid advancement of artificial intelligence (AI) is driving demand for high-capacity semiconductor memories to store increasingly large model weights and intermediate data, including key-value (KV) caches during inference \cite{Review_IEDM}. Meeting this demand requires continued improvements in memory density and cost per bit. However, as conventional DRAM advances toward 10-nm-class technology nodes, maintaining sufficient cell capacitance within shrinking footprints, while controlling leakage and sustaining data retention, presents increasingly significant scaling challenges. 

These challenges motivate alternative device architectures and cell layouts that increase storage density beyond conventional planar implementations. In particular, advances in fabrication have enabled one-transistor-one-capacitor (1T-1C) memory cells to evolve from conventional 6F$^2$ layouts toward compact 4F$^2$ architectures based on vertical-channel transistors (VCTs), where \textit{F} denotes the minimum feature size \cite{DRAM_1T1C}. Fig.\ref{fig:intro}(a) illustrates 4F$^2$ memory cell structures based on storage and sensing mechanism.

Beyond the growing demand for memory capacity, the deployment of AI in space exploration and orbital computing platforms imposes stringent requirements on memory reliability under radiation exposure \cite{s1,s2,s3}. In such environments, energetic particles, particularly heavy ions, can strike semiconductor devices, generating a dense plasma of electron--hole pairs along their trajectories. These strikes can induce single-event effects (SEEs), potentially disturbing stored states and causing bit flips, referred to as single-event upsets (SEUs) \cite{space_radiation}. Such errors can corrupt critical information, including model weights and intermediate computational data, thereby compromising the accuracy and reliability of AI workloads \cite{s4}. This vulnerability is particularly concerning for emerging 4F$^2$ memory structures. Unlike conventional 6F$^2$ cells employing recessed-channel transistors, emerging 4F$^2$ architectures adopt vertically aligned source–channel–drain configurations, often without a dedicated body contact \cite{DRAM_1T1C}, making them susceptible to radiation-induced floating-body effects as illustrated in Fig.~\ref{fig:intro}(b). Radiation resilience is, therefore, a critical consideration in assessing the viability of emerging high-density memory technologies for spaceborne AI applications.

Although prior studies have investigated radiation-induced disturbances in semiconductor memories \cite{space_radiation, zhao2015single}, including ferroelectric memory cells \cite{Wodzro}, systematic comparisons across cell and array architectures with fundamentally different storage mechanisms remain underexplored, particularly when accounting for array biasing, parasitic loading, and sensing circuitry. Understanding how these factors jointly determine radiation susceptibility is necessary to identify vulnerabilities and guide the design of reliable, high-density memory systems.

Beyond transistor footprint, capacitor geometry also constrains memory density. Conventional 1T-1C DRAM uses a metal–insulator–metal (MIM) dielectric capacitor to provide sufficient storage capacitance, imposing substantial structural and fabrication overhead as cell dimensions shrink. Capacitorless 2T-0C gain cells offer an alternative by storing data as charge on the effective storage-node (SN) capacitor and sensing the resulting voltage through a separate read transistor. Wide-bandgap oxide-semiconductor (e.g. Indium–gallium–zinc–oxide (IGZO)) channels can suppress off-state leakage, while their compatibility with low-temperature processing facilitates multitier integration. Eliminating the explicit storage capacitor therefore offers a path toward compact, vertically integrated memories \cite{2T0C_high_den}.

However, the reduced SN capacitance also introduces a radiation-reliability trade-off. A heavy-ion strike deposits energy along its trajectory, generating electron–hole pairs in the semiconductor. The resulting carrier collection and transient increase in transistor conduction can disturb the stored SN charge. For a given voltage margin, the critical charge required to cause an upset scales with SN capacitance. Consequently, the small effective SN capacitance of 2T-0C cells can make them more susceptible than 1T-1C DRAM to strike-induced voltage disturbances and subsequent readout errors, despite their low standby leakage.

This vulnerability motivates storage mechanisms that reduce reliance on retained free charge. Ferroelectric RAM (FeRAM) replaces the dielectric storage capacitor with a metal–ferroelectric–metal (MFM) capacitor, encoding data in two remanent polarization states ($\pm$ \textit{P}\textsubscript{R}). Although polarization-based storage is less susceptible to transient charge loss, access transistors remain vulnerable to ion strikes, which can induce sufficient voltage transients across the ferroelectric capacitor to disturb its polarization. Such disturbances may be further influenced by array biasing and neighboring-cell operations.

This work examines how SN capacitance influences readout reliability under heavy-ion strikes, evaluates the density–radiation-resilience trade-offs of 4F$^2$ volatile-memory architectures, and investigates the benefits and limitations of ferroelectric polarization storage as a mitigation approach. The analysis incorporates array biasing, parasitic loading, and sensing circuitry to connect device-level disturbances with readout outcomes. The main contributions are as follows:

\begin{itemize}
    \item We identify that any hold-state drain--source bias as a key condition for heavy-ion-induced storage disturbance.
    \item We show that reduced storage-node capacitance in capacitorless DRAM creates a trade-off between memory density and radiation resilience.
    \item We demonstrate FeRAM improves resilience due to its inherent properties, data encoding, and storage mechanism; however, we also reveal that operation on a selected row can compromise FeRAM resilience in half-selected columns by enabling polarization disturbance.
    \item We demonstrate that multiple capacitor stacking can lead to multi-bit polarization disturbance or flips threatening system reliability.
\end{itemize}

\newlength{\tcadrowheight}

\begin{figure*}[t]
    \centering
   
        \centering
       
            \includegraphics[width=0.9\linewidth]
                {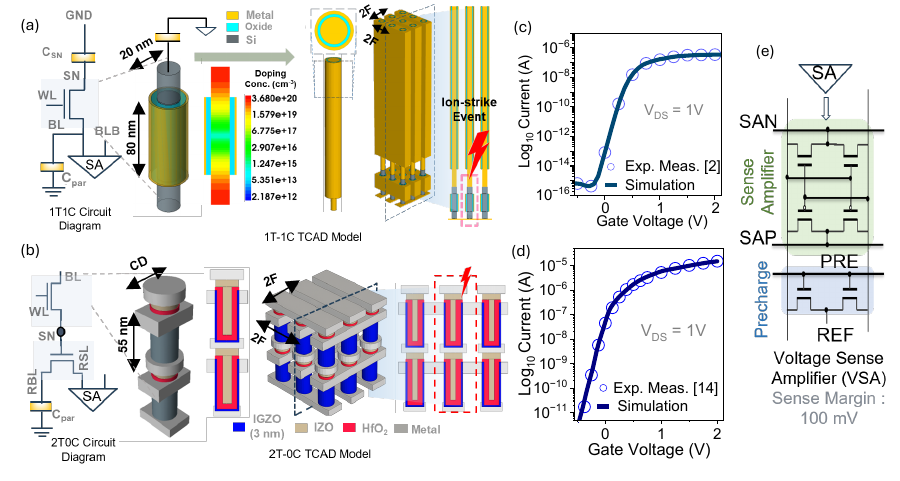}%

    \caption{%
TCAD modeling and experimental calibration of 4F² DRAM architectures. Circuit schematics, device structures, array configurations, and heavy-ion strike locations for (a) 1T–1C and (b)  2T–0C DRAM, respectively. Calibration of simulated $I_D$–$V_G$ characteristics against experimental measurements for (c) 1T-1C and (d) 2T-0C access transistor. (e) Voltage sense amplifier used in mixed-mode simulations.}
    \label{fig:TCAD}
    \vspace{-5mm}
\end{figure*}

\begin{figure}[t]
    \centering
        \includegraphics[width=0.8\columnwidth]
            {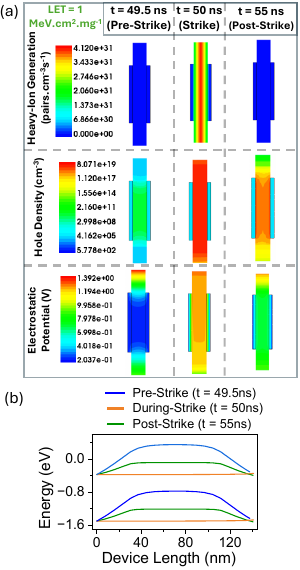}%

    \caption{
    Heavy-ion-induced carrier dynamics and energy barrier modulation at an LET of 1 MeV·cm²/mg. (a) Spatial evolution of carrier generation, hole density, and electrostatic potential. (b) Energy band profiles before, during, and after the strike.}
    \label{fig:example_1_MeV_Band_Diagram}

    \vspace{-3mm}
\end{figure}

\begin{figure}[t]
    \centering
        \includegraphics[width=\columnwidth]
            {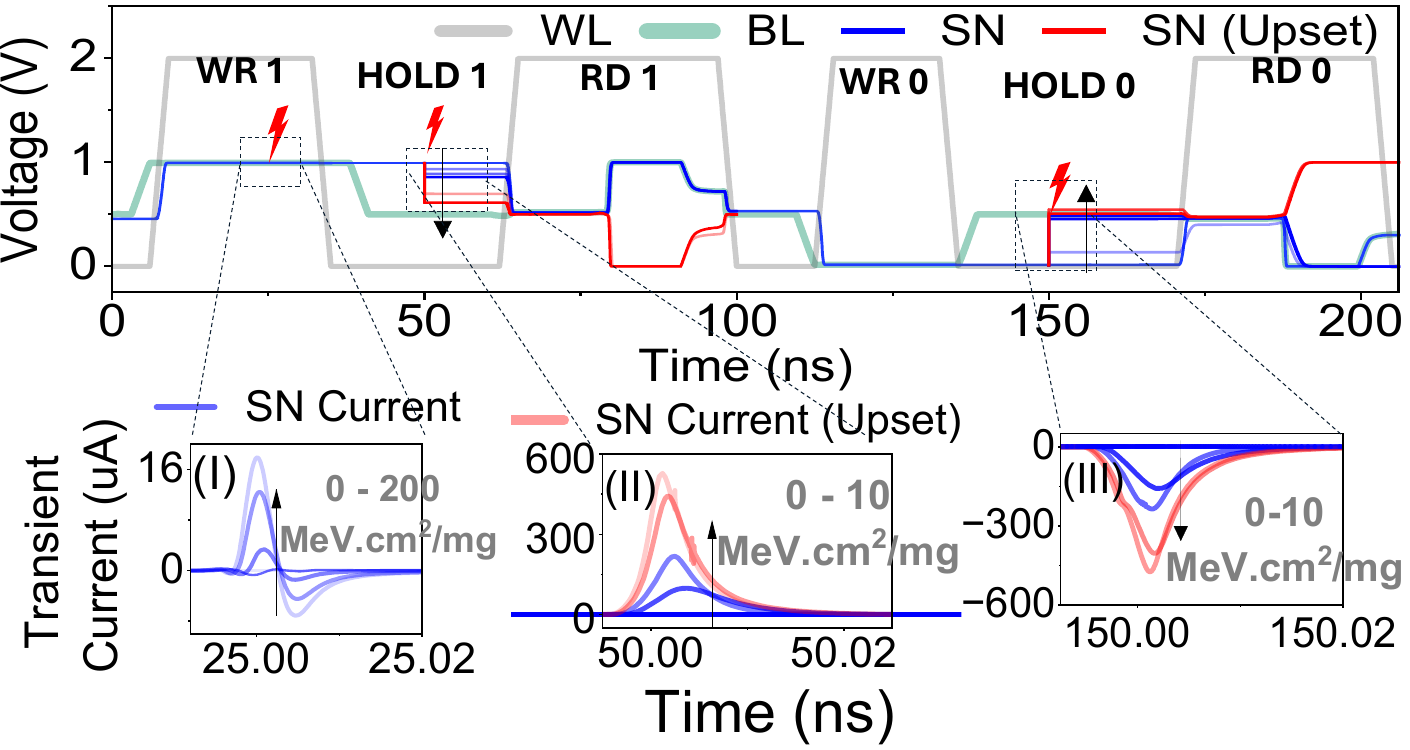}%

    \caption{Transient 1T-1C DRAM operation under heavy-ion strikes during write and hold, illustrating LET-dependent SN current transients (insets) and resulting storage-node voltage disturbances and readout errors in both '1'/'0' states.}
    \label{fig:1t1c_mechanism}
\end{figure}

\begin{figure}[t]
    \centering
    
\includegraphics[width=\columnwidth]
            {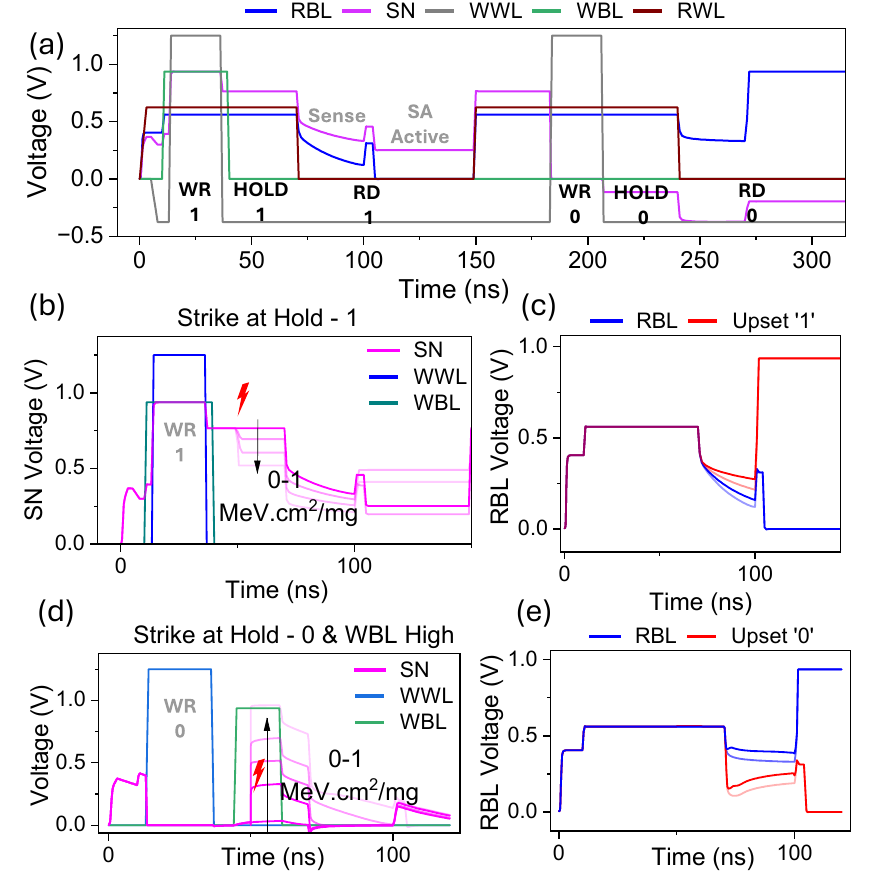}%

    \caption{(a) 2T-0C DRAM operation,
    (b) ion-strike during hold-1 state causing SN voltage deviation, and (c) RBL readout disturb/upset,  during hold-0, when shared WBL rises to write 1 to other cell, (d) ion-strike deviates stored SN leading to (e) RBL disturb/upset.}
    \label{fig:2t0c_mechanism}

    \vspace{-4mm}
\end{figure}

\begin{figure}[t]
     \centering
    \vspace{-4mm}

 \includegraphics[width=\linewidth]{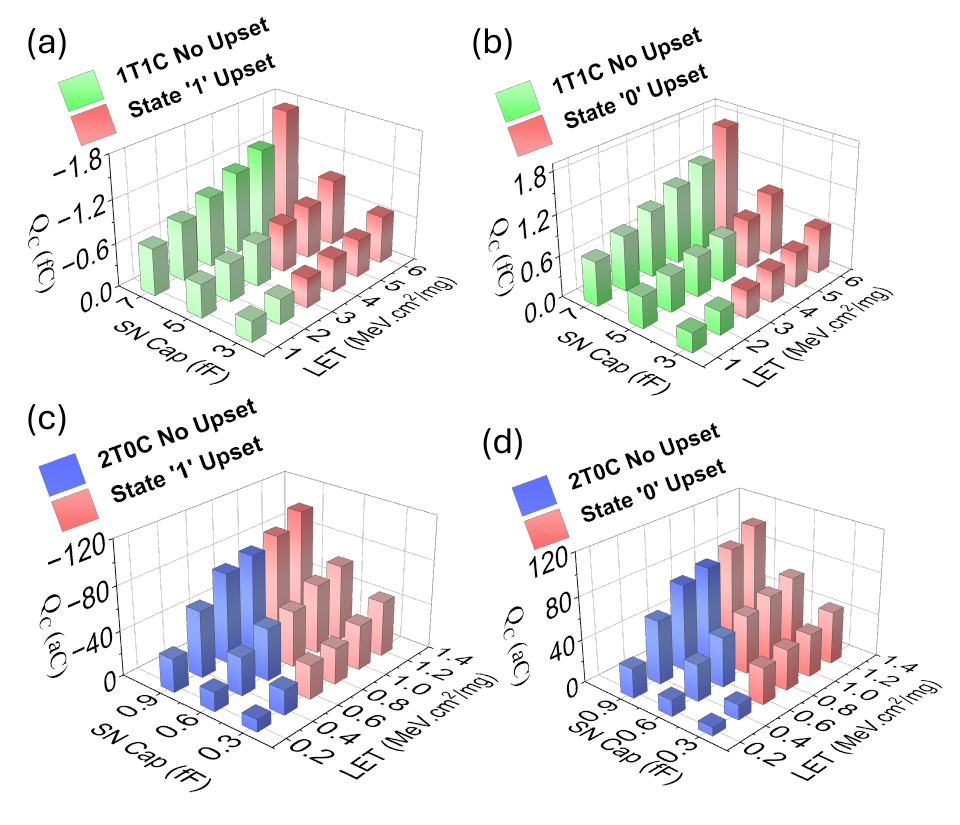}%

    \caption{During a heavy-ion strike in the hold-`1' state,
    a nonzero SN--BL potential difference can disturb the stored
    state and cause a bit flip in (a)-(b) 1T--1C DRAM despite its high SN
    capacitance. The lower effective SN capacitance of (c)-(d) 2T--0C DRAM
    results in an approximately $10\times$ lower LET threshold.}
    \label{fig:3D_Plots_1t1c_2t0c}

    \vspace{-2 mm}
        
\end{figure}

\begin{figure}[t]
    \centering
    \vspace{-4mm}
    \subfloat[]{%
        \includegraphics[width=0.5\linewidth,valign=c]
        {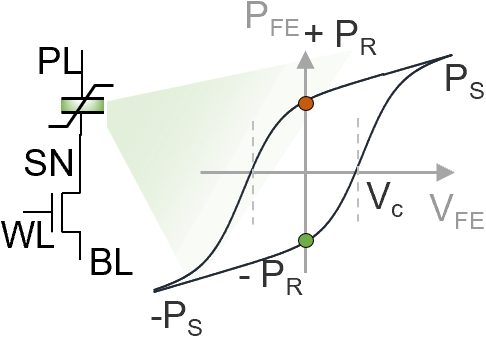}%
        }%
     \hfill
    \subfloat[]{%
        \includegraphics[width=0.48\linewidth,valign=c]
        {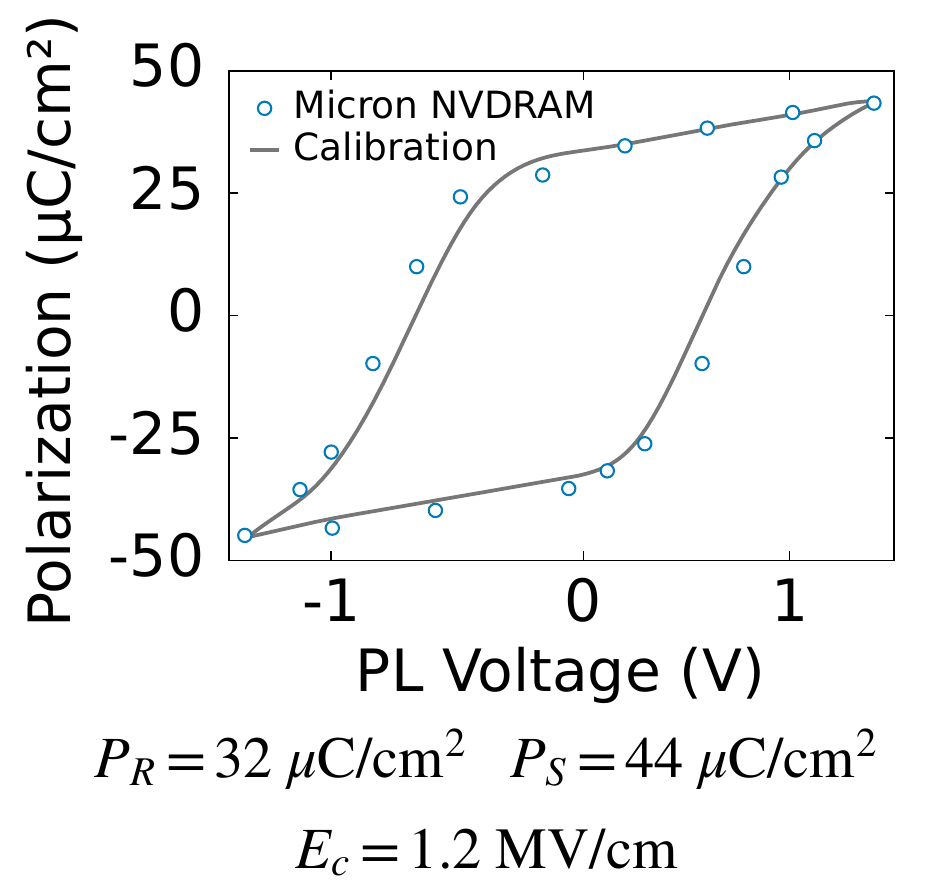}%
        }%
    \caption{(a) 1T-1C FeRAM has similar structure to 1T-1C DRAM and utilizes MFM capacitance with bistable remanent polarization ($+P_R$ \& $-P_R$), (b) In this work, the P-V loop is calibrated to experimental measurements \cite{NVDRAM}}
    \label{fig:FeRAM_diagram_PV}
    \vspace{-0mm}
\end{figure}

\begin{figure}[t]
    \centering
    \vspace{-4mm}
        \includegraphics[width=\linewidth,valign=c]
        {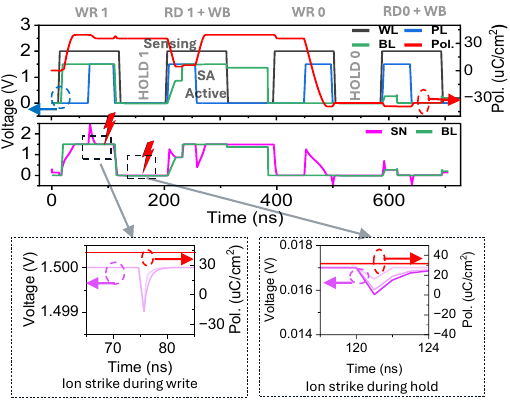}%
    \caption{Transient FeRAM write, hold, and read simulation with write-back, showing negligible polarization disturbance despite strike-induced SN voltage transients during write and hold.}
    \label{fig:FeRAM_target_mechanismm}
    \vspace{-5mm}
\end{figure}

\section{Cell Architecture Driving Reliability Trade-off in Volatile 1T-1C and 2T-0C DRAM}

DRAM has traditionally relied on 1T–1C cells for their density, speed, and cost. Increasing capacity demands have driven the transition from 6F$^2$ layouts to vertical-channel-transistor (VCT)-based 4F$^2$ cells. However, maintaining storage capacitance requires high-aspect-ratio MIM capacitors, increasing fabrication complexity and cost. Capacitorless 2T–0C DRAM with IGZO channel instead stores charge on the effective SN capacitance for gain-cell readout. However, reduced SN capacitance increases sensitivity to charge-induced voltage disturbances, with radiation susceptibility further influenced by charge collection, device characteristics, biasing, and sensing margins.
This section investigates heavy-ion-induced single-event effects in 4F$^2$ 1T–1C and 2T–0C DRAM. Potential extensions to multilayer 3D DRAM are left for future work.

\subsection{Design Framework for TCAD Simulation}
To capture the electrical and radiation responses of 1T--1C VCT DRAM, we adopted the junctionless gate-all-around (GAA) architecture reported in~\cite{DRAM_1T1C} to model in Sentaurus TCAD. Fig.~\ref{fig:TCAD}(a) presents the 1T-1C circuit diagram with added parasitics and peripherals, three-dimensional (3D) cell, and $3 \times 3$ array structure of 1T--1C DRAM model. A metal–insulator–metal (MIM) capacitor is connected to the storage node, with its dimensions adjusted to achieve capacitances of $3$, $5$, and $7~\mathrm{fF}$. These variations enable evaluation of the impact of storage-node capacitance on radiation resilience. The model has been extended as $3\times3$ array to capture parasitic effects from neighbors and shared interconnects at the center cell. The $I_D - V_G$ characteristic of the access transistor has been calibrated to the experimental data from \cite{DRAM_1T1C}, shown in Fig.~\ref{fig:TCAD}(c).

The 2T--0C model in Fig.~\ref{fig:TCAD}(b), adopted from \cite{2T_0C_base}, employs vertically stacked channel-all-around (CAA) IGZO transistors. Unlike conventional 1T--1C DRAM, 2T--0C cells require separate write and gain-based read transistors to compensate for limited SN capacitance and minimize leakage. Therefore, a practical 2T--0C architecture with comparable cell dimensions, rather than identical 1T-1C transistor structures, is adopted to enable a representative architecture-level radiation comparison. IGZO was modeled in Sentaurus TCAD as a custom oxide semiconductor with a 3.2-eV bandgap, two exponential band-tail distributions, and two Gaussian donor/acceptor trap distributions \cite{IGZO_Channel}. The $I_D$--$V_G$ characteristics were calibrated to experimental measurements \cite{2t0c_exp}, as shown in Fig.~\ref{fig:TCAD}(d).

To capture the device response to radiation exposure, physical models were selected to account for intrinsic carrier transport and strike-induced carrier dynamics. Lombardi mobility and high-field velocity saturation models were employed to capture carrier transport, while Shockley--Read--Hall (SRH) and Auger recombination models accounted for excess carrier recombination. Hurkx band-to-band tunneling accounted for field-induced tunneling leakage, and the Okuto avalanche model was incorporated to capture impact-ionization-induced carrier multiplication. Heavy-ion strikes were modeled using Gaussian radial and temporal carrier-generation profiles, with a characteristic radius of $10$ nm \cite{char_radius_1,strike_ref_new} and a track length of $1$ $\mu$m spanning the cell along the strike direction.

To evaluate how device-level disturbances translate into functional memory errors, both device models were integrated with external readout circuitry through Sentaurus mixed-mode transient simulations. Read- and write-line parasitic capacitances of $20$ fF were incorporated to capture interconnect loading and its influence on transient responses. A cross-coupled voltage sense amplifier \cite{sheikholeslami2000survey} [Fig.~\ref{fig:TCAD}(e)], implemented using 32-nm low-power PTM transistors \cite{32nm_PTM}, enables reliable readout with a sensing margin of $100$ mV. %

\vspace{-2mm}
\subsection{SEE in 1T-1C DRAM Architecture}
Fig.~\ref{fig:1t1c_mechanism} illustrates the 1T--1C DRAM operating sequence, including write (WR), read (RD), and hold states for both logic levels. During write, the asserted wordline (WL) connects the SN to the bitline (BL), charging or discharging the SN toward the applied BL voltage. During readout, charge sharing between the SN and precharged BL is followed by sensing and restoration. In contrast, during hold, the WL is deasserted, isolating the SN while the BL remains at $V_{DD}/2$. The impact of a heavy-ion strike strongly depends on the operating state, particularly the drain-to-source bias ($V_{DS}$) across transistor.

An energetic heavy ion generates a dense electron--hole plasma along its trajectory. To evaluate device performance in the worst-case scenario, we study ion-strike traversing the longest path along the center and parallel to the transistor length. In nanoscale Si channels, strong local electric fields, short transport distances, and high electron mobility facilitate rapid electron collection toward the source/ drain contacts. In contrast, lower hole mobility, limited hole-collection pathways, and the absence of a dedicated body contact can obstruct hole removal, leading to transient positive space-charge accumulation and a floating-body effect (FBE). Fig.~\ref{fig:example_1_MeV_Band_Diagram}(a) demonstrates the carrier-generation rate, hole density, and channel electrostatic potential before, during, and after a strike at an LET of $1~\mathrm{MeV\cdot cm^2/mg}$. The accumulated positive charge raises the channel electrostatic potential, lowering the conduction-band barrier for electron injection and transient conduction. 

Under nominal hold conditions, the energy barrier remains sufficiently high to suppress carrier transport between the drain (BL) and source (SN). However, the strike-induced accumulation of holes elevates the body potential, substantially lowering the energy barrier and establishing a transient conduction path despite the deactivated gate (WL). This loss of electrostatic isolation facilitates charge transfer between the BL and SN, worsened by non-zero $V_{DS}$, potentially disturbing the stored state. Following the strike, the barrier gradually recovers but remains partially suppressed due to residual hole accumulation. This radiation-induced barrier modulation is illustrated in Fig.~\ref{fig:example_1_MeV_Band_Diagram}(b), which captures the energy barrier evolution before, during, and after the strike at $t=50$ ns.

During active write and restoration, the SN is driven toward the intended BL voltage ($V_{DS} = 0$) and limit voltage-driven charge transfer during strike-induced conduction. Accordingly, (I) in Fig.~\ref{fig:1t1c_mechanism} shows negligible disturbance during write-`1', even at LET values up to $200~\mathrm{MeV\cdot cm^2/mg}$. Similar behavior is observed in the other examined read/write conditions.

During hold, however, the deasserted WL isolates the SN while the BL remains at $V_{DD}/2$, establishing a nonzero $V_{DS}$. Under this bias, strike-induced barrier lowering transiently enhances off-state conduction, allowing charge transfer between the SN and BL and driving the SN voltage toward the standby BL potential. (II) and (III) in Fig.~\ref{fig:1t1c_mechanism} demonstrate this behavior during hold-`1' and hold-`0', respectively. Increasing LET produces larger SN disturbances, potentially causing readout upsets at  $\simeq3$--$6~\mathrm{MeV\cdot cm^2/mg}$, depending on storage.

\subsection{SEE in 2T-0C DRAM Architecture}
\par

The capacitorless 2T--0C architecture employs gain-cell readout to sense data stored at the SN as shown in Fig.\ref{fig:TCAD}(b). During writing, the write wordline (WWL) activates the write transistor, allowing the write bitline (WBL) to set the SN voltage to a high or low level. The SN serves as the gate of the read transistor, whose conductance determines the discharge rate of the read bitline (RBL). The resulting RBL voltage is sensed by a sense amplifier, as shown in Fig.~\ref{fig:2t0c_mechanism}(a).

The write transistor plays a particularly important role in the single-event response because it directly controls charge transfer between the WBL and SN. The TCAD-extracted storage-node (SN) capacitances of 2T--0C DRAM are substantially smaller than those of conventional 1T--1C DRAM, measuring $0.3$, $0.6$, and $0.9$ fF at critical dimensions (CDs) of $60$, $90$, and $120$ nm, respectively. The small effective SN capacitance stores substantially less charge than the dedicated capacitor in 1T--1C DRAM at the same voltage. Consequently, a comparable charge loss produces a larger SN-voltage deviation, potentially exceeding the threshold and causing an upset.

Similar to 1T--1C DRAM, a heavy-ion strike during writing has minimal impact on the stored state once the SN and WBL potentials equalize and $V_{DS} \approx 0$.

As shown in Fig.~\ref{fig:2t0c_mechanism}(b), during hold-`1', the SN remains high, while the WWL and WBL return to $0$V, establishing a nonzero $V_{DS}$ across the write transistor. A heavy-ion strike at any instance during hold, allows charge transfer between the SN and WBL and reduces the SN potential. The resulting disturbance can cause an incorrect RBL readout [Fig.~\ref{fig:2t0c_mechanism}(c)] with upsets observed at LET values below $1~\mathrm{MeV \cdot cm^{2}/mg}$.

During hold-'0', WWL, WBL, and SN remain near 0V, suppressing strike-induced disturbances. However, writing '1' to another cell sharing the WBL raises its voltage while the unselected WWL remains low, establishing a nonzero $V_{DS}$ across the write transistor [Fig.~\ref{fig:2t0c_mechanism}(d)]. Under such bias, a heavy-ion strike triggers transient conduction and raise the SN potential, potentially upsetting the stored `0' during subsequent RBL readout [Fig.~\ref{fig:2t0c_mechanism}(e)].

\begin{figure*}[t]
    \centering
   
        \centering
       
            \includegraphics[width=\linewidth]
                {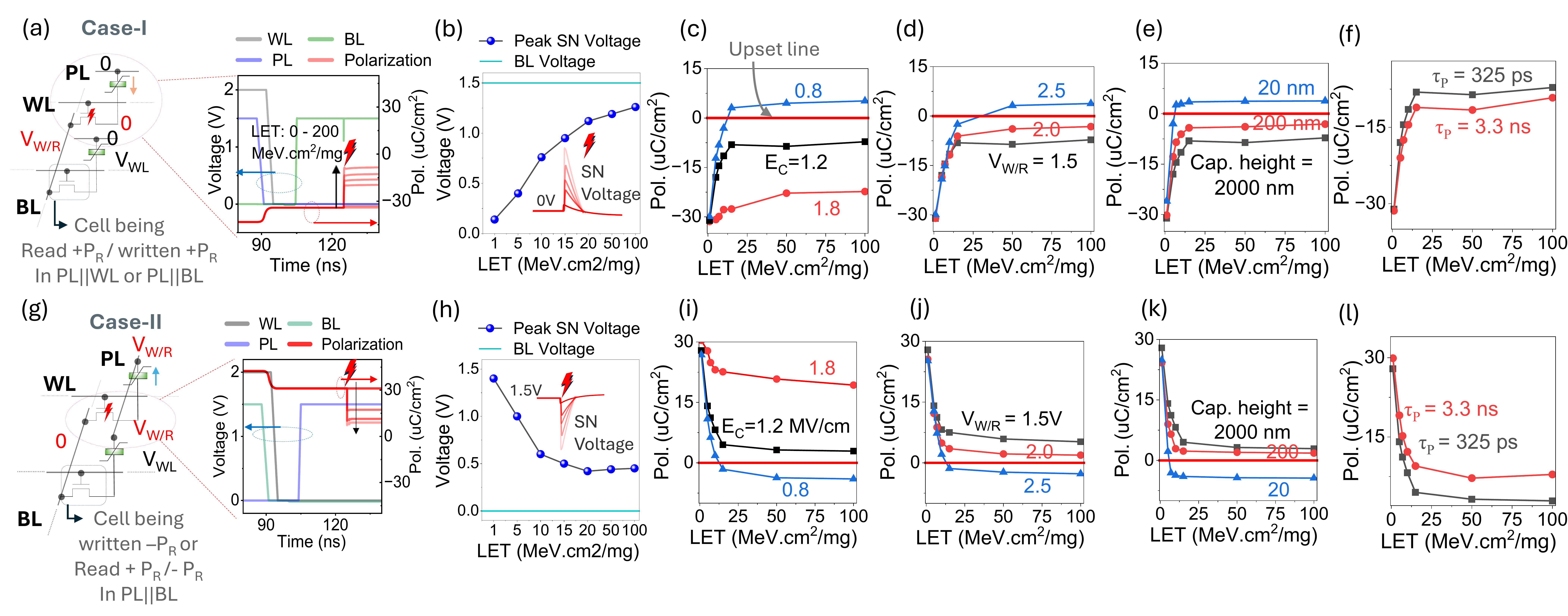}%

    \caption{%
        Radiation-induced polarization disturbance in FeRAM under two vulnerable bias conditions. (a), (g) Case-I and Case-II bias configurations and transient polarization responses. (b), (h) LET-dependent peak SN voltage. Impact of (c), (i) coercive field, (d), (j) write voltage, (e), (k) FeCAP height, and (f), (l) polarization relaxation time on radiation-induced polarization degradation}
    \label{fig:Pol_Disturb}
    \vspace{-5mm}
\end{figure*}

\subsection{Density and SEE Resilience Trade-Off In 1T-1C and 2T-0C DRAM}
Although 2T--0C DRAM enables high-density, low-temperature 3D integration with reduced fabrication complexity by eliminating dedicated MIM capacitors, its smaller SN capacitance increases susceptibility to ion-induced soft errors compared with conventional 1T--1C DRAM. Figs.~\ref{fig:3D_Plots_1t1c_2t0c}(a) and~\ref{fig:3D_Plots_1t1c_2t0c}(b) show the collected charge at the SN ($Q_C$) as a function of LET for different 1T--1C storage capacitances during hold-`1' and hold-`0', respectively. Increasing the SN capacitance from 3 to 7 ${fF}$ raises the critical charge required for an upset from approximately 0.6 to 1.8 $fC$, with the corresponding upset LET increasing from approximately 3 to $6~\mathrm{MeV \cdot cm^{2}/mg}$.

In contrast, 2T--0C DRAM exhibits substantially smaller effective SN capacitance of $0.3$--$0.9$ fF, with critical charges increasing from $40$ to $120$ aC  and upset LET thresholds ranging from $0.2$ to $1$ MeV$\cdot$cm$^2$/mg [Fig.~\ref{fig:3D_Plots_1t1c_2t0c}(c)--(d)]. These thresholds are approximately an order of magnitude lower than those of 1T--1C DRAM, reflecting the combined influence of smaller SN capacitance and differences in transistor characteristics and charge collection. Thus, the comparison highlights the radiation-reliability trade-offs between practical $4F^2$ memory architectures rather than isolating capacitance effects alone.

\section{Impact of Storage Medium on Soft Errors: Volatile vs. Nonvolatile Ferroelectric}
Hafnia-based 4F$^{2}$ FeRAM are attractive candidates for dense nonvolatile storage owing to their CMOS compatibility, scalability, and advancing technological maturity. Unlike DRAM, FeRAM stores data in two stable remanent polarization ($+P_R$ \& $-P_R$) states that persist without an externally applied bias. Modeling and experimental studies have demonstrated strong radiation resilience in individual FeRAM cells \cite{Wodzro}. However, their resilience also depends on the bias conditions imposed by the memory architecture within an array operation. In this section, we investigate the radiation resilience and potential vulnerabilities of FeRAM cells under array-level operating conditions across array configurations.

\subsection{Modeling FeRAM in TCAD}
The primary structural difference between 1T--1C DRAM and 1T--1C FeRAM is the replacement of the MIM capacitor with an HZO-based metal-ferroelectric-metal (MFM) capacitor. For a fair comparison, the same 1T-1C device model with similar dielectric capacitance are used, with ferroelectric polarization enabled HZO in the ferroelectric capacitor. Ferroelectric polarization was modeled using the TCAD Preisach model, with the simulated polarization-voltage (P-V) characteristics calibrated with experimental measurements reported in~\cite{NVDRAM} as shown in Fig.~\ref{fig:FeRAM_diagram_PV}(b). A polarization relaxation time of $325$ ps and $3$ ns, estimated using the $3\tau$ approximation of experimentally reported switching times \cite{325psref, NVDRAM} to understand the role of switching time.

\subsection{SEE Resilience in Selected FeRAM Cell}
In FeRAM the logic states ``1'' and ``0'' are encoded by the two remanent polarization states, $+P_{\mathrm{R}}$ and $-P_{\mathrm{R}}$, respectively as shown in Fig.~\ref{fig:FeRAM_diagram_PV}(a). Fig.~\ref{fig:FeRAM_target_mechanismm} shows the simulated waveforms of write/read states 1/0 in 1T-1C FeRAM having the work mechanism compliant with \cite{FeRAM_GATech, NVDRAM}. During write/read stage, access transistor is activated to create necessary voltage difference across the ferroelectric capacitor (FeCAP) to encode the polarization state. In contrast to 1T-1C DRAM operation, during hold state, both SN and BL are at low potential ($0V$), thus when WL is deactivated, $V_{DS}$ is normally always at $0V$.

To disturb the stored state, an ion strike must generate a sufficiently large electric field opposite to the current polarization state stored in FeCAP to shift it from one stable state toward the other. Fig.~\ref{fig:FeRAM_target_mechanismm} demonstrates two cases of heavy-ion irradiation (0-200 $MeV \cdot cm^2/mg$) during write and hold $+P_R$ stage to demonstrate the strike effects. During write stage, as the WL is already active and BL is shorted to SN, ion-strike does not alter the SN potential. Again, during hold state, although WL is off, as $V_{DS}$$=$$0V$, ion-strike only causes negligible transient change in the SN potential. Thus, the target cell is highly resilient to irradiation under normal conditions.

\subsection{Polarization Disturbance in Half-Selected FeRAM}
Although FeRAM exhibits strong resilience under zero-bias hold conditions, half-selected cells can become susceptible to heavy-ion-induced disturbances during array operations. The first vulnerability arises when reading/ writing $+P_R$ state from/ to a cell in the selected row, which requires the BL to rise to $V_{W/R}$. Consequently, half-selected cells sharing the same BL experience a nonzero $V_{DS}$ across their access transistors, as their WLs remain deasserted and SNs at 0V. Under this condition, an ion-strike can transiently activate the access transistor and disturb the stored $-P_R$ polarization referred to as  Case~I, illustrated in [Fig.~\ref{fig:Pol_Disturb}(a)]. This vulnerability exists in both PL$\parallel$WL and PL$\parallel$BL architectures, where only BL is shared across columns.

An additional disturbance mechanism arises specifically in PL$\parallel$BL architectures where PL and BL both are shared across columns. During any read or $-P_R$ writing where the BL remains low while the shared PL rises to $V_{W/R}$. In half-selected rows, the WL remains deasserted, leaving the SN electrically isolated from the BL. Capacitive coupling from the shared PL raises the SN potential, resulting in a negligible voltage across the FeCAP but a nonzero $V_{DS}$ across the access transistor. A subsequent ion strike can induce transient conduction and alter the voltage across the FeCAP, potentially disturbing the stored $+P_R$ polarization, as shown in Case~II [Fig.~\ref{fig:Pol_Disturb}(f)]. This mechanism is absent in PL$\parallel$WL architectures because the PL bias is confined to the selected row.

In both scenarios, with increasing ion-strike energy a transient spike at SN with peak voltage closer to BL voltage is observed that disturb the stored polarization states [Fig.~\ref{fig:Pol_Disturb}(b), (h)]. Therefore, coercive field ($E_C$), read/write bias ($V_{W/R}$), FeCAP dimension and switching time ($\tau_P$) are critical design parameters governing polarization stability and SEE resilience under half-selected conditions.

 Increasing $E_C$ from 0.8 to 1.8 MV/cm raises the critical LET by nearly an order of magnitude, demonstrating that a higher polarization switching barrier effectively suppresses strike-induced polarization reversal [Fig.~\ref{fig:Pol_Disturb}(c), (i)]. Again, raising $V_{W/R}$ from 1.5 to 2.5 V reduces the critical LET approximately 15–30 MeV·cm²/mg [Fig.~\ref{fig:Pol_Disturb}(d), (j)], suggesting that stronger write bias establishes a more vulnerable electrostatic condition during radiation exposure. Reducing the capacitor height from 2000 to 20 nm lowers the critical LET [Fig.~\ref{fig:Pol_Disturb}(e), (k)], highlighting a fundamental trade-off between vertical scaling and radiation resilience, as reduced capacitance amplifies the voltage disturbance for a given collected charge. Increasing $\tau_P$ by an order of magnitude improves the residual polarization by approximately 3–5 µC/cm² [Fig.~\ref{fig:Pol_Disturb}(e), (k)] at the highest LET, indicating that slower polarization dynamics provide intrinsic temporal filtering of polarization disturbance at the expense of write latency.

\begin{figure}[t]
    \centering

    \includegraphics[width=0.485\textwidth]{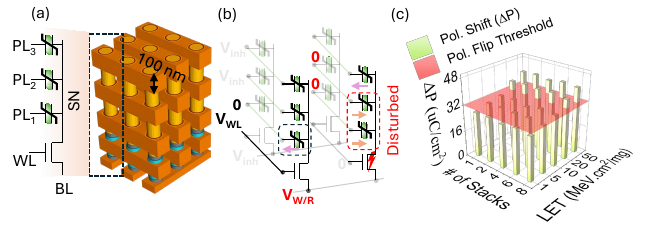}%
        
    \caption{(a) TCAD model of a $3 \times 3$ xT-nC FeRAM, 
(b) multi-bit disturb mechanism, (c) Worst-case analysis showing post ion-strike polarization trend with stack and LET.}
    \label{fig:multi_bit_flips}

    \vspace{-5mm}
\end{figure}

\vspace{-4mm}
\subsection{Multi-Bit Disturb In xT-nC FeRAM Cells}
Unlike DRAM, FeRAM stores data through remanent polarization and therefore does not require a sustained SN voltage for data retention. This nonvolatile storage mechanism enables vertically stacked FeCAPs to share a common SN, increasing memory density while reducing the transistor overhead per bit leading to multi-capacitor structures such as 1T-nC \cite{intel_1tnc}, which is natural extension to 1T-1C and 2T-nC \cite{2T_nC_first_demo} with gain-cell which utilizes partial switching. Fig.~\ref{fig:multi_bit_flips} presents the common -$n$C architecture with multiple vertically stacked FeCAP layers with reduced FE capacitor height (100 nm).

Similar to the previously discussed cases I \& II, an unselected xT--$n$C cell may contain multiple FeCAPs storing the $-P_{\mathrm{R}}$ state while a neighboring-cell operation drives the shared BL high. A common xT-nC structure is presented in Fig.~\ref{fig:multi_bit_flips}(a). Under nonzero $V_{\mathrm{DS}}$ condition, such as Fig.~\ref{fig:multi_bit_flips}(b), heavy-ion-induced leakage can perturb the common SN voltage and simultaneously disturb the polarization states of the connected FeCAPs. A worst case analysis with xT-nC FeRAM with all FeCAPs storing negative $P_R$ is irradiated with heavy-ion. Fig.~\ref{fig:multi_bit_flips} (c) plots the average polarization disturbance across all FeCAPs with increasing stack counts and heavy-ion energy. Polarization upset occurs at critical $\mathrm{LET}\leq 15~\mathrm{MeV}\cdot\mathrm{cm}^{2}/\mathrm{mg}$ with similar disturbance at all FeCAPs which can potentially deteriorate system performance if the cell contains important data such as, most-significant-bit of model weights.

\section{Conclusion}
We reveal the density--reliability trade-offs of $4F^{2}$ volatile and nonvolatile memories under heavy-ion strikes. The low SN capacitance of 2T--0C DRAM improves density but reduces its LET threshold by approximately $10\times$ compared with 1T--1C DRAM. FeRAM provides greater resilience through polarization-based storage, although biased unselected cells remain vulnerable. Optimizing device parameters, biasing conditions, and FeCAP stacking can further suppress ion-strike-induced disturbances.

\bibliographystyle{IEEEtran}

\bibliography{references_updated}

\end{document}